# MINIJET PRODUCTION AT SMALL $x$

A Donnachie
Department of Theoretical Physics, University of Manchester
and
P V Landshoff
DAMTP, University of Cambridge

**Abstract**

Soft pomeron exchange at high energy includes minijet production at central rapidity values. This predicts copious production of minijets in deep inelastic scattering at small $x$, which is closely related to the fact that $\nu W_2$ exhibits Regge behaviour. It helps to explain also how the integrated inclusive cross-section for minijet production in $\gamma p$ or $pp$ collisions can be much greater than the total cross-section.

**1 Introduction**

Regge theory successfully predicted[1] the $\gamma p$ total cross-section measured[2] at HERA. There were some other predictions that were much larger[3], and it is interesting to discuss why they turned out to be wrong. They were based on perturbative calculations of inclusive minijet production. Minijets are jets of rather small transverse momentum, say less than 1% of the total available energy $\sqrt{s}$. While their transverse momentum $q_T$ has to be large enough for perturbation theory to be used to predict their production rate, it may be too small to allow a clean experimental analysis. Since the scale of nonperturbative effects is typically 1 GeV, we would expect that lowest-order perturbation theory reproduces the inclusive cross-section $d\sigma/dq_T$ reasonably successfully down to $q_T^{\min} \approx 1$ GeV, with a $K$-factor somewhere between $\frac{1}{2}$ and 2. One calculates $d\sigma/dq_T$ from the familiar diagram of figure 1, which involves the structure functions of the incoming particles and a hard scattering and yields a pair of minijets of approximately equal and opposite transverse momenta. If we integrate it, we obtain[4]

$$I_1 = \frac{1}{\sigma^{\text{TOT}}} \int_{q_T^{\min}} dq_T \, \frac{d\sigma^{\text{pair}}}{dq_T} = \bar{n} \, \rho(q_T^{\min}) \tag{1.1}$$

where $\rho(q_T^{\min})$ is the fraction of events containing minijets, and $\bar{n}$ is the average number of minijet pairs in these events. The naive expectation from figure 1 is that exactly one minijet pair is produced, that is $\bar{n} = 1$ in (1.1). However, one finds if one evaluates the integral in (1.1) and compares with the HERA measurements, that this would require $\rho(q_T^{\min}) > 1$, which is wrong by definition.

So it must be that $\bar{n} > 1$. This is not really a surprise. One way in which $\bar{n}$ can be greater than 1 is from multiple parton-parton scatterings[5] involving two or more partons from each of the initial particles. We would expect this to be very important in high-energy nucleus-nucleus reactions[6], but we doubt whether it is the main mechanism in $\gamma p$ or $pp$ collisions.

Rather, there is another mechanism[4], which is intrinsically nonperturbative. Further, while maybe the effects of multiple parton collisons can be approximately handled by an eikonalisation procedure, eikonalisation is not relevant for this other mechanism, which may be understood as follows. In order to calculate $d\sigma/dq_T$ from figure 1, one needs the two structure functions down to fractional-momentum values $x$ of order $x \sim q_T^{\min}/\sqrt{s}$, that is very small values when $\sqrt{s}$ is large. But, from elementary kinematics, when a parton of momentum $k$ is pulled out of a particle of momentum $p$ with $k = xp + \ldots$, the squared invariant mass of the system it leaves behind is[7]

$$s_0 \sim -\frac{k^2 + \mathbf{k}_T^2}{x} \qquad (1.2)$$

and so is large when $x$ is small. That is, the upper and lower clusters of residual fragments of the initial hadrons in figure 1 each have large invariant mass and so they are very likely to contain additional minijets in some fraction of the events.

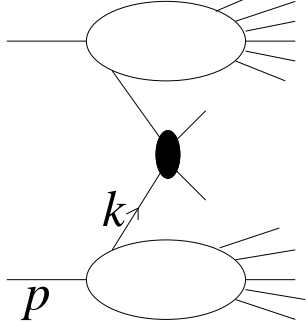

Figure 1: Diagram for single-inclusive minijet-pair production

Of course, the structure function is measured directly in deep inelastic lepton scattering, where also the residual fragments of the proton will have large invariant mass at small $x$ and so are likely to include minijets. These minijets will be distributed uniformly in rapidity in the central region. In this paper we calculate this effect. We expect it to be of significant magnitude, because it is closely connected with the reason for structure functions exhibiting Regge behaviour at small $x$.

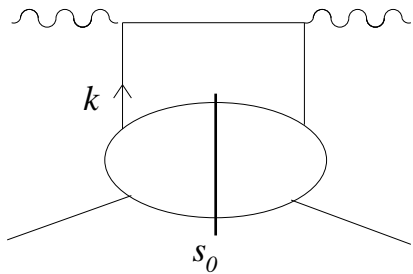

Figure 2: Parton model for $\nu W_2$

Consider $\nu W_2$ at a value of $Q^2$ that is only moderately large, so that perturbative evolution has not yet set in and the parton model applies: figure 2. The squared energy of the lower bubble, which is the amplitude for finding a parton $k$ in the proton $p$, again satisfies (1.2). Since this bubble is an elastic strong-interaction amplitude, its high-energy behaviour is governed by Regge theory and is just a sum of terms $s_0^{\alpha(0)}$. If we insert this into the calculation of $\nu W_2$ from figure 2, we obtain[7] a sum of terms $(1/x)^{\alpha(0)-1}$. NMC data[8] at moderate $Q^2$ and not-too-small $x$ show that $\nu W_2$ contains such Regge terms at small $x$, approximately a constant from soft-pomeron exchange and close to $\sqrt{x}$ from $f_2, a_2$ exchange[9].

It is not widely known that such a behaviour, which we stress is nonperturbative, arises[1] because the proton fragments that remain when a small-$x$ parton is pulled out have large invariant mass. We note that the Monte Carlos used at HERA set both $k^2$ and $k_T$ to zero (before the perturbative evolution begins) and so do not make $s_0$ large. Thus they miss a nonperturbative effect which is surely important.

The occasional production of high-$p_T$ jets is part of the Regge behaviour; in particular, it is part of soft pomeron exchange. As the energy increases, this minijet production becomes more and more frequent, but it is still part of soft pomeron exchange and does not change the effective power of $s$ with which the total cross section rises. We calculate the probability that soft pomeron exchange includes such production.

In section 2, we introduce the coupling of the soft pomeron to a gluon and estimate its strength from the small-$x$ behaviour of the proton's gluon distribution. We use this in section 3 to estimate minijet production in deep inelastic lepton scattering at small $x$, and find that it should be copious for small $q_T^{\min}$.

## 2 Inclusive cross-section for production of a single pair of minijets

As is well known[10], at small $x$ there is no $k_T$ ordering, so that perturbative and nonperturbative effects are inevitably intermingled. In particular, as the dominant mechanism for the production of two pairs of minijets in hadron-hadron collisions, we propose figure 3. As we go down the diagram, soft pomerons alternate with hard interactions that produce the minijets. If we slice the diagram down the middle, the left half corresponds to figure 1, with the upper (or lower) hard interaction in figure 3 corresponding to the one shown explicitly in figure 1, and the other one being implicitly included within the lower (or upper) bubble in figure 1. Gluonic partons are more efficient at producing high-$p_T$ jets than are quarks[11], so we concentrate on the case where the hard collisions are glue-glue.

This means that we need the coupling of the soft pomeron to a gluon. In order to extract this, we first calculate the inclusive cross-section for production of a single pair of minijets. That is, we calculate figure 4. For the coupling of the pomeron to an on-shell quark we have used previously[1]

$$\beta_0 \gamma^\alpha \qquad (2.1a)$$

together with a signature factor that ensures that quarks and antiquarks couple with equal sign. This is very successful phenomenologically, with

$$\beta_0 \approx 2 \text{ GeV}^{-1} \qquad (2.2a)$$

We need the coupling of the soft pomeron to a gluon only at zero momentum transfer $t$. At $t = 0$ the most general coupling to a gluon of momentum $k$ would be

$$\delta^{ab} A(k^2) \, g^{\mu\nu} \, k^\alpha \qquad (2.1b)$$

together with terms in $k^\mu$ or $k^\nu$ (or both), where $\mu$ and $\nu$ are Lorentz indices for the gluon and $a$ and $b$ are the corresponding colour indices. To construct a high-energy forward scattering amplitude for either $qq$, $qg$ or $gg$ scattering, we multiply a corresponding pair of couplings (2.1) by $(\alpha' s_0)^\epsilon$, where $s_0$ is the squared centre-of-mass energy for the amplitude and $\epsilon$ and $\alpha'$ are the intercept and slope of the pomeron trajectory:

$$\epsilon \approx 0.08 \qquad \alpha' = 0.25 \text{ GeV}^{-2} \qquad (2.2b)$$

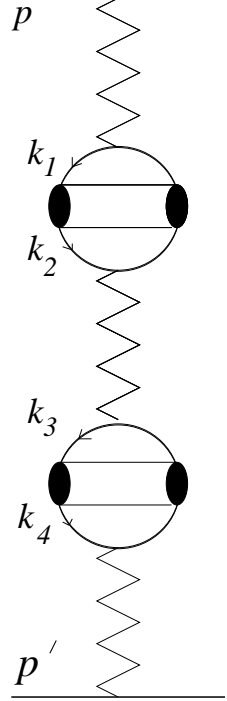

Figure 3: Diagram for double-inclusive minijet-pair production, with alternating soft pomeron exchange and hard scattering

When the gluon is on-shell, only the term in its coupling shown explicitly in (2.1b) contributes, as the others vanish when multiplied by the gluon polarisation vectors. In most applications where the gluon is not on shell, it is still true that we need only this term, because the other ends of the gluon lines are coupled to an amplitude that is annihilated by $k^\mu$ or $k^\nu$. We include in $A(k^2)$ a (nonperturbative) propagator for each gluon leg. As we shall see, we do not need to know the function $A(k^2)$, but only the value of the integral

$$g = \frac{\alpha'^\epsilon}{(2\pi)^4}(2+\epsilon)^{-1} \int_{-\infty}^0 dk^2 (-k^2)^{2+\epsilon} A(k^2) \tag{2.3}$$

The calculation of figure 4 involves the square $M$ of the hard-scattering $gg$ amplitude, integrated over the momenta of the final-state jets – or rather, because of momentum conservation, over the momentum $q$ of just one of them. Because of the $g^{\mu\nu}\delta^{ab}$ in the coupling (2.1b) of the pomeron to a gluon, the Lorentz indices and colour indices for $M$ are summed in the same way as in the calculation of the differential cross-section $d\sigma/d\hat{t}$ for the sub-process. In fact,

$$M = \frac{512\hat{s}^2}{\pi} \int \frac{d^3q}{q^0} \delta(\hat{s}+\hat{t}+\hat{u}) \frac{d\sigma}{d\hat{t}} \tag{2.4}$$

To calculate figure 4 we introduce momenta $k_1$ and $k_2$ as indicated, and express each in the form

$$k_i = x_i p - y_i p' + k_{iT} \tag{2.5}$$

In our calculation, we may neglect the masses of the particles $p$ and $p'$. Although there is no $k_T$ ordering, there is $x$ and $y$ ordering[10] in the region of phase space that dominates in the calculation:

$$1 \gg x_1 \gg x_2 \gg 0, \qquad 0 \ll y_1 \ll y_2 \ll 1 \tag{2.6}$$

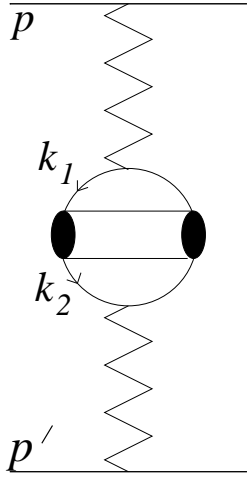

Figure 4: Diagram for single-inclusive minijet-pair production, with hard scattering between soft pomeron exchanges

What we mean by these strong inequalities is that if we write $x_1 \sim (\alpha' s)^{-\eta_1}$ and $x_2 \sim (\alpha' s)^{-\eta_2}$ then $0 < \eta_1 < \eta_2$. We use

$$\int d^4k_1 \sim \tfrac{1}{2}\pi \int \frac{dx_1}{x_1} dk_1^2 \, d\mathbf{k}_{1T}^2 \qquad \int d^4k_2 \sim \tfrac{1}{2}\pi \int \frac{dy_2}{y_2} dk_2^2 \, d\mathbf{k}_{2T}^2 \qquad (2.7a)$$

The subenergies across the two pomerons satisfy

$$s_1 \sim y_1 s \sim -\frac{k_1^2 + \mathbf{k}_{1T}^2}{x_1} \qquad s_2 \sim x_2 s \sim -\frac{k_2^2 + \mathbf{k}_{2T}^2}{y_2} \qquad (2.7b)$$

The subenergy $s_1$ is raised to the power $\epsilon$ and is multiplied by the coupling of the gluon to the pomeron, so that we encounter the integral

$$g = \frac{\alpha'^\epsilon}{(2\pi)^4} \int_{-\infty}^0 dk_1^2 \int_0^{-k_1^2} d\mathbf{k}_{1T}^2 (-k_1^2 - \mathbf{k}_{1T}^2)^{1+\epsilon} A(k_1^2) \qquad (2.7c)$$

where the upper limit on the $k_{1T}$ integration arises from the need to keep $s_1$ positive. This is the only way in which the gluon coupling enters the calculation; on performing the $k_{1T}$ integration we retrieve (2.3).

The coupling function $A(k^2)$ of the gluon to the soft pomeron is nonperturbative, and we have defined it to include also propagators for the two gluon legs. Thus we expect[12] that it is appreciable in size only when $|k^2|$ is less than about 1 GeV$^2$. We see from (2.7c) that there is a similar constraint on $\mathbf{k}_T^2$, so the partons $k_1$ and $k_2$ have only rather small transverse momentum and the transverse momenta of the two minijets produced in the hard collison are almost equal and opposite. So, knowing that each jet is paired with a second one in this way, we calculate the inclusive cross cross section for producing a pair of minijets:

$$q^0 \frac{d\sigma^{\text{pair}}}{d^3q} = \frac{1}{2} \int dx_1 dy_2 F(x_1) F(y_2) \delta(s_{12} + t_{12} + u_{12}) \frac{s_{12}}{\pi} \frac{d\sigma}{dt_{12}} \qquad (2.8a)$$

with $s_{12}, t_{12}$ and $u_{12}$ the Mandelstam variables for the central hard scattering (so that $s_{12} = (k_1 - k_2)^2 \sim x_1 y_2 s$), and

$$F(x) = 24\pi\beta_0 \ gx^{-1-\epsilon} \tag{2.9}$$

From the familiar structure of (2.8a), we identify $F(x)$ as the gluon structure function of the proton at small $x$. From its known normalisation[13], we deduce that for $Q^2 \approx 5 \text{ GeV}^2$,

$$g \approx 15 \text{ MeV} \tag{2.10}$$

If we integrate over all $q_T > q_T^{\min}$, $M$ in (2.4) becomes

$$M(\hat{s}) = 512\hat{s} \int_{t_-}^{t_+} d\hat{t} \frac{d\sigma}{d\hat{t}} \theta(\hat{s} - 4q_T^{\min 2}), \qquad t_\pm = -\tfrac{1}{2}\hat{s} \pm \tfrac{1}{2}\sqrt{\hat{s}(\hat{s} - 4q_T^{\min 2})} \tag{2.11}$$

and the integrated inclusive cross-section for producing a pair of minijets is

$$\int_{q_T^{\min}} dq_T \frac{d\sigma^{\text{pair}}}{dq_T} = \frac{9\pi^2 \beta_0^2 g^2}{16s} \int dx_1 dy_2 (x_1 y_2)^{-2-\epsilon} M(x_1 y_2 s) \tag{2.12}$$

We change integration variables from $y_2$ to $s_{12} = x_1 y_2 s$. We then perform the $x_1$ integration, subject to (2.6), which has become

$$1 \gg x_1 \gg \frac{s_{12}}{s} \tag{2.13}$$

We also obtain from (2.6) the conditions $s_{12} \gg -(k_i^2 + \mathbf{k}_{iT}^2)$ ($i = 1, 2$), but these can be ignored because (2.11) requires $s_{12} > 4q_T^{\min 2}$, and the form factor that couples the soft pomeron to the gluon suppresses contributions to the integral (2.3) from large $(k_i^2 + \mathbf{k}_{iT}^2)$. Because the limits (2.13) on the $x_1$ integration only enter in a logarithm, at high energy it is a good approximation to replace the strong conditions (2.13) by simple inequalities. Thus, with $\sigma^{TOT} = 18\beta_0^2(\alpha's)^\epsilon$, (2.12) becomes

$$I_1 = \frac{1}{\sigma^{TOT}} \int_{q_T^{\min}} dq_T \frac{d\sigma^{\text{pair}}}{dq_T} = \frac{\pi^2 g^2}{32\alpha'^\epsilon} \int_{4q_T^{\min 2}} ds_{12} s_{12}^{-2-\epsilon} M(s_{12}) \log\left(\frac{s}{s_{12}}\right) \tag{2.14}$$

where $M(s_{12})$ is given in (2.11).

Although we have explicitly considered $pp$ scattering, the result (2.14) is valid for the scattering of any pair of hadrons. It is valid also when one (or both) of the hadrons is replaced with a real photon. This is because, in minijet production, as in the total cross section, the photon behaves like a hadron: the "direct" component of its structure function is unimportant[14].

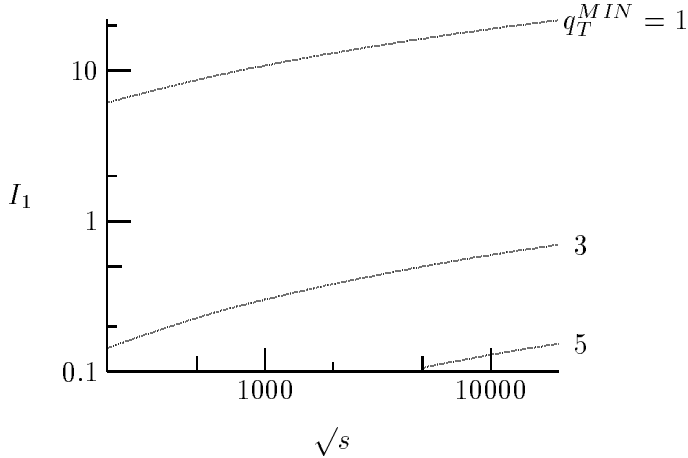

Figure 5: The integral $I_1$ of (2.14)

We are unable to calculate (2.14) analytically. Numerical computation, taking account only of $gg \to gg$ in the hard scattering, gives the results shown in figure 5. We have used a running coupling $\alpha_s$ evaluated at $4q_T^{\min^2}$ and with $\Lambda_{QCD} = 250$ MeV, and introduced a $K$-factor of 2 to take account of non-leading-order effects. We recall that the integrated inclusive cross section has the interpretation given in (1.1), so that it is no surprise that it is larger than the total cross-section.

As a check on our calculation, we have compared our results with the UA1 minijet data[15] in figure 6. In this experiment the jet event cross section was defined as the cross section for producing at least one cluster with $E_T^{\text{raw}} >5$ GeV at any rapidity. A nominal 5 GeV cut on the cluster transverse energy is quoted as corresponding to an effective average parton $p_T$ threshold of between 3 and 4 GeV and this effective threshold depends, in an unspecified way, upon the c.m. energy. We have chosen to compare our results with the data using a $p_T$ cut of 4 GeV, and as can be seen from figure 6 the comparison is reasonable. As our calculation is asymptotic, in the sense that we have taken the $x \to 0$ limit (our structure function is simply $x^{-\epsilon}$, with no accompanying power of $(1-x)$), our estimate of the inclusive minijet cross section is less meaningful at the lower energies than at the higher.

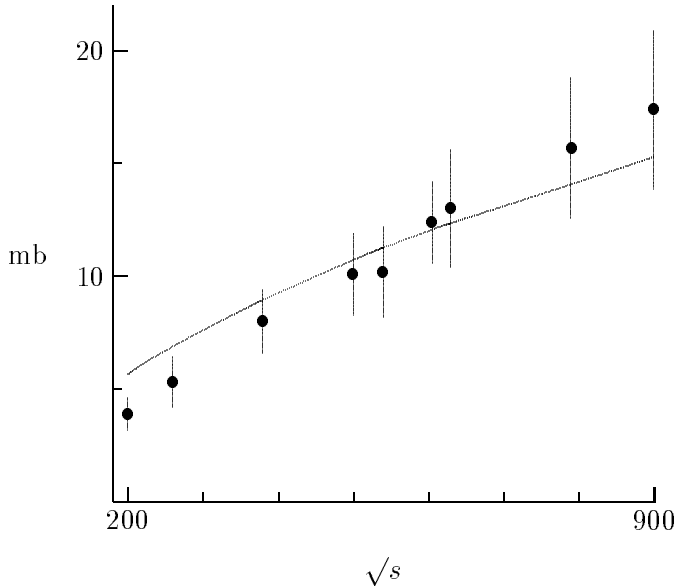

Figure 6: Comparison of calculated inclusive minijet cross-section integrated down to $q_T = 4$ GeV with UA1 data[15]

### 3. Deep inelastic lepton scattering at small $x$

We now calculate how much minijet production may be expected in the central rapidity region in deep inelastic scattering.

Pomeron exchange is supposed to dominate[7] in the small-$x$ behaviour of $\nu W_2$. At small and moderate values of $Q^2$ it is expected that the pomeron that is involved is the soft pomeron, so that

$$\nu W_2 \sim C(Q^2) x^{-\epsilon} \tag{3.1}$$

with $\epsilon \approx 0.08$ as is given in (2.2b). This expectation is well verified[9] by NMC data[8] at moderate $Q^2$ and not-too-small $x$, though the measurement at $Q^2 = 8.5$ GeV$^2$ by the H1 collaboration[16] gives

Even if the effective value of $\epsilon$ increases with $Q^2$, the effect of this should to some extent cancel in the ratio of the minijet cross-section to the total cross-section (see (3.3) below). In any case, whatever mechanism is responsible for the rise in the effective value of $\epsilon$ will surely increase minijet production. So we here confine our attention to the contribution from soft pomeron exchange, and interpret our results as lower limits for the fraction of events that contain minijets. We assume that soft pomeron exchange factorises, so that (3.1) corresponds to figure 7a. The probability that the soft pomeron exchange includes the production of a pair of minijets is then calculated from figure 7b. As before, we expect that the main mechanism for minijet production is gluon-initiated, and confine ourselves to considering this.

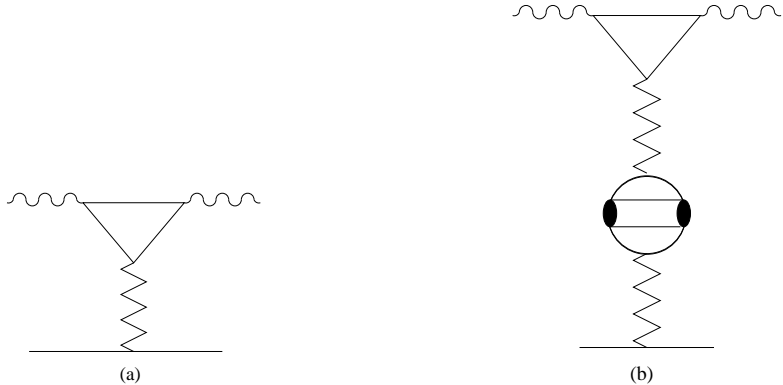

Figure 7: (a) Parton model for $\nu W_2$ at small $x$ 16 (b) Single-inclusive minijet-pair production in deep inelastic scattering at small $x$

The calculation is similar to that outlined in section 2, with $p'$ replaced with $Q$ and the condition on $x_2$ in (2.6) strengthened so that $x_2 \gg x$. This has the consequence that (2.13) becomes

$$1 \gg x_1 \gg \frac{x s_{12}}{\mu^2} \qquad (3.2)$$

where $\mu^2 = -(\mathbf{k}_2^2 + \mathbf{k}_{2T}^2)$. Strictly, the appearance of $\mu^2$ here prevents us from factoring off the $\mathbf{k}_2^2$ and $\mathbf{k}_{2T}^2$ integrations and absorbing them into the integral (2.7c) that defines $g$, but at very large $s$ the error in this is small. This is because, as we have explained, the pomeron's coupling to the gluon, together with the nonperturbative gluon propagators, effectively constrain $\mu$ to be of the order of 1 GeV or so, and since it will appear only in a logarithm we can set it constant without too much error. We find that, for $Q^2 > \mu^2$, the fraction of events in which there is a pair of minijets with $q_T > q_T^{\min}$ is

$$\frac{1}{\nu W_2(x, Q^2)} \int_{q_T^{\min}} dq_T \frac{d}{dq_T} \nu W_2 = \frac{\pi^2 g^2}{32 \alpha'\epsilon} \int_{4 q_T^{\min 2}}^{\mu^2/x} ds_{12} \, s_{12}^{-2-\epsilon} M(s_{12}) \log\left(\frac{\mu^2}{x s_{12}}\right) \qquad (3.3)$$

One might have thought that the argument of the logarithm here should be $Q^2/x s_{12}$, that is $s/s_{12}$, in conformity with (2.14). At extremely large values of $s$ (or extremely small values of $x$) the two are equivalent, but at subasymptotic values the kinematics lead to the form in (3.3).

We plot the fraction (3.3) in figure 8 for two values of $q_T^{\min}$. The bands in the figure correspond to $\mu^2$ varying in the range 0.5 to 2 GeV$^2$. Note that $q_T^{\min}$ refers to the parton: contributions from the underlying event background must be subtracted from data.

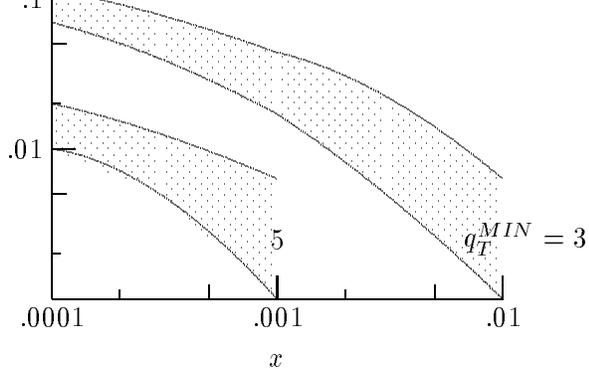

Figure 8: Fraction of deep-inelastic-scattering events that contain minijets

## 4 Production of two pairs of minijets in $pp$ or $\gamma p$ collisions

The calculation of figure 3 is along the same lines. It yields the inclusive cross section for the production of two pairs of minijets. The transverse momenta of the minijets in each pair are almost equal and opposite, and the total-momentum vectors of the two pairs are randomly distributed in rapidity in the central region of rapidity, and therefore, since we assume that the total energy is very high and a large rapidity interval is available, they are usually well separated. We express each of the four $k_i$ in the form (2.5), and obtain the dominant contribution from $x$ and $y$ ordering:

$$1 \gg x_1 \gg x_2 \gg x_3 \gg x_4 > 0 \qquad 0 \ll y_1 \ll y_2 \ll y_3 \ll y_4 \ll 1 \qquad (4.1)$$

The integrated inclusive cross-section for producing two pairs of minijets is

$$I = \int_{q_T^{\min}} dq_T dq'_T \frac{d^2 \sigma^{\text{pairs}}}{dq_T dq'_T} = \frac{9\pi^4 \beta_0^2 g^4}{256 \alpha'^{\epsilon}} \int dx_1 dy_2 dx_3 dy_4 (x_1 y_2 x_3 y_4 s)^{-2-\epsilon} M(x_1 y_2 s) M(x_3 y_4 s) \qquad (4.2)$$

We change integration variables from $y_2$ to $s_{12} = x_1 y_2 s$ and from $x_3$ to $s_{34} = x_3 y_4 s$. We then perform the $x_1$ and $y_4$ integrations, subject to the conditions (4.1). The important conditions read

$$y_2 \gg s_{12}/s \qquad x_3 \gg s_{34}/s \qquad y_2 x_3 s \ll \mu^2 \qquad (4.3a)$$

where now

$$\mu^2 = \min\left(-k_2^2 - \mathbf{k}_{2T}^2, -k_3^2 - \mathbf{k}_{3T}^2\right) \qquad (4.3b)$$

With these conditions,

$$\int \frac{dx_1}{x_1} \int \frac{dx_3}{x_3} \sim \tfrac{1}{2} \log^2 R \ \theta(R-1) \qquad (4.4a)$$

with

$$R = \frac{\mu^2 s}{s_{12} s_{34}} \qquad (4.4b)$$

As we have explained, the appearance of $\mu^2$ prevents us from factoring off the $k_2^2, k_3^2, \mathbf{k}_{2T}^2$ and $\mathbf{k}_{3T}^2$ integrations and absorbing them into the integral (2.7c) that defines $g$, At extremely large $s$ the error in doing this is small. The result of setting $\mu^2$ constant is that (4.2) becomes

$$I_2 = \frac{1}{\sigma^{\text{TOT}}} \int_{q_T^{\min}} dq_T dq'_T \frac{d^2 \sigma^{\text{pairs}}}{dq_T dq'_T}$$

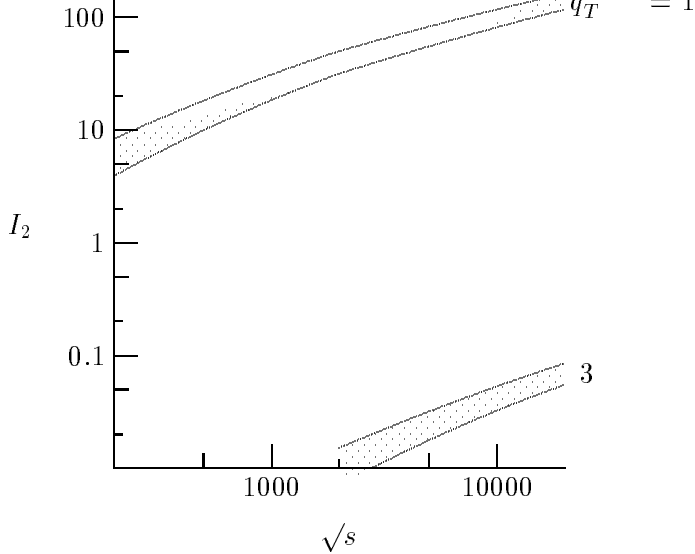

Figure 9: The integral $I_2$ of (4.5)

$$= \frac{\pi^4 g^4}{1024\alpha'^{2\epsilon}} \int_{4q_T^{\min 2}} ds_{12} ds_{34}\, (s_{12}s_{34})^{-2-\epsilon} M(s_{12}) M(s_{34}) \log^2 R\ \theta(R-1) \qquad (4.5)$$

In figure 9 we plot this for various values of $q_T^{\min}$. The bands again correspond to $\mu^2$ in the range 0.5 to 2 GeV$^2$.

The interpretation of the doubly-integrated inclusive cross section 3.2) is analogous to (1.1):

$$I_2 \leq \bar{n}(\bar{n}-1)\rho(q_T^{\min}) \qquad (4.6)$$

where again $\rho(q_T^{\min})$ is the fraction of events that contain at least one minijet pair, and $\bar{n}$ is the average number of pairs of minijets (including the pair containing the "trigger" minijet) in those events. We have written (4.6) as an inequality, in order to account of the other mechanisms for providing an inclusive cross-section for two minijet pairs, for example those involving more than one parton from each initial particle[5]. The need for the presence of some other mechanism is evident because, from its definition, $\rho(q_T^{\min}) \leq 1$, so that it must be that

$$\frac{I_1^2}{I_1 + I_2} \leq 1 \qquad (4.7)$$

## 5 Discussion

In this paper we have considered inclusive cross-sections for minijet production, and have emphasised that they are not related in a simple way to total cross sections. We have found that there is a nonperturbative mechanism, which forms part of soft pomeron exchange and is closely related to the small-$x$ behaviour of structure functions, that for small $q_T^{\min}$ provides copious minijet production in high-energy hadron-hadron, photon-hadron and photon-photon collisions, and in deep inelastic lepton scattering at small $x$. The mechanism is not included in the standard Monte Carlos, for example those used to analyse HERA data, and we urge that this be put right.

The mechanism is equally applicable to the production of heavy flavour in the central region, and yields charm and beauty cross sections in high energy hadron-hadron interactions that are compatible with data. As is the case for the minijets, this does not affect the overall rise in the total cross section.

The nonperturbative mechanism of course includes perturbative hard-scattering sub-processes. We have included only glue-glue subprocesses, but for minijet production the consequent under-estimation will be very small. We have also worked only to lowest order in the perturbative $\alpha_S$, evaluated at $4q_T^{\min^2}$ with $\Lambda_{QCD} = 250$ MeV. We have rather arbitrarily chosen a $K$-factor of 2, and there is also the uncertainty on the value of $g$, mainly arising from the poorly-known gluon structure function at very small $x$. Nonetheless for the inclusive minijet cross section the order of magnitude is sensible, as is shown by the comparison in figure 6 with UA1 data.

The calculations show that it is incorrect to ascribe the rise in the photon-hadron or hadron-hadron total cross section to the onset of minijet production. The minijets can be seen to be an integral part of the pomeron, which is itself the controlling mechanism for the total cross section.

*This research is supported in part by the EC Programme "Human Capital and Mobility" Network "Physics at High Energy Colliders" contract CHRX-CT93-0537 (DG 12 COMA)*